\documentclass[doublecol]{epl2}
\usepackage{amsmath,amssymb,graphicx}
\usepackage{bm}% bold math
\definecolor{labelkey}{cmyk}{.4,.2,0,0}
\renewcommand{\epsilon}{\varepsilon}
\usepackage{graphicx,color}
\graphicspath{{./}{./figures/}{./Figures/}}
\usepackage{dcolumn}% Align table columns on decimal point
\usepackage{bm}% bold math
\usepackage{dsfont}

\newcommand{\rme}{\mathrm{e}}

\newcommand{\rmd}{{\mathrm{d}}}
\newcommand{\nn}{\nonumber}
\newcommand{\fig}[2]{{\includegraphics[width=#1]{#2}}}
\newcommand{\Fig}[1]{{\includegraphics[width=8.6cm]{#1}}}

\def\be{\begin{equation}}
\def\ee{\end{equation}}

\def\bea{\begin{eqnarray}}
\def\eea{\end{eqnarray}}

\begin{document}

\title{Non-Gaussian effects and multifractality in the Bragg glass}

\author{Andrei A. Fedorenko$^{1}$, Pierre Le Doussal$^{2}$ and Kay J\"org Wiese$^{2}$}
\institute{$^{1}$CNRS-Laboratoire de Physique, Ecole Normale Sup\'erieure de Lyon, 46 all\'ee  d'Italie, 69007\ Lyon, France.\\
$^{2}$CNRS-Laboratoire
de Physique Th{\'e}orique de l'Ecole Normale Sup{\'e}rieure, 24 rue
Lhomond,75005 Paris, France.}
%\date{\today}
\shortauthor{A. A. Fedorenko, P. Le Doussal and K. J. Wiese}
\date{\today\ -- \jobname}%\ -- compilation \input{\jobname.counter}}

%\pacs{68.35.Rh}{Phase transitions and critical phenomena}

\abstract{We study, beyond the Gaussian approximation, the
decay of the translational order correlation function for a $d$-dimensional
scalar periodic elastic system in a disordered environment.
We develop a method based on functional determinants,
equivalent to summing an infinite set of diagrams. We obtain, in dimension  $d=4-\epsilon$, the
even \(n\)-th cumulant of relative displacements as  $\overline{\left<[u(r)-u(0)]^n\right>^{\rm c} } \simeq {\cal A}_n  \ln r$ with ${\cal A}_n = -
 { (\varepsilon/3)^{n} \Gamma(n-\frac12) \zeta(2n-3)
  }/{  \sqrt{\pi} }$,
  %where \(a\) is a microscopic cutoff
  as well as the multifractal dimension $x_q$ of the exponential field $e^{q u(r)}$. As a corollary, we obtain an analytic expression for a class of $n$-loop integrals in $d=4$, which appear in the perturbative determination of Konishi amplitudes, also accessible via AdS/CFT  using integrability.}

\date{\today}
\maketitle

%%%%%%%%%%%%%%%%%  Introduction %%%%%%%%%%%%%%%%%%%%%%%%

{\it Introduction:} Periodic elastic systems in quenched disorder model numerous applications, from charge-density waves in solids~\cite{Gruner1988},
vortex lattices in superconductors~\cite{BlatterFeigelmanGeshkenbeinLarkinVinokur1994,GiamarchiLeDoussalBookYoung
%,LeDoussal2010Book
}
%,BognerEmigNatterman2001},
Wigner crystals \cite{AndreiDevilleGlattliWilliamsParisEtienne1988}, Josephson junction arrays \cite{GranatoKosterlitz1989},
to liquid crystals~\cite{RadzihovskyToner1999}.
%In all these systems
The competition between elastic energy, which favors periodicity, and disorder,
which favors distortions, produces a complicated energy landscape
with many metastable states. While we know since Larkin \cite{Larkin1970}
that weak disorder destroys perfect translational order, it was
realized later that topological order (i.e.\ no dislocations)
may survive, leading to the Bragg glass phase (BrG)
\cite{GiamarchiLeDoussal1994,
%GiamarchiLeDoussal1995,
GiamarchiLeDoussalBookYoung
%LeDoussal2010Book
} and validating the elastic
description. A key observable, measured from the structure factor
in diffraction experiments \cite{KleinJoumardBlanchardMarcusCubittGiamarchiLeDoussal2001}, is the translational
correlation function $C_K({\bf r})=\langle\overline{e^{iK[u(\mathbf{r})-u(0)]}}\rangle$,
where $u(\mathbf{r})$ is the ($N$-component) displacement of a node from its position in the
perfect lattice, and $K$ is chosen as a reciprocal lattice vector (RLV). Overlines stand for disorder averages,
and brackets for thermal averages. Thermal fluctuations  are subdominant,
and we focus  on $T=0$.
It was established \cite{Nattermann1990,GiamarchiLeDoussal1994
%,GiamarchiLeDoussal1995
} that
at large scale $u(\mathbf{r})$ is {\it a log-correlated field},
\bea \label{logc}
\overline{\left<[u({\bf r})-u(0)]^2\right>} \simeq {\cal A}_2 \ln \frac{r}{a}
,\eea
where $a$ is a microscopic cutoff, and $r:=|{\bf r}|$. If one further assumes $u({\bf r})$ to be Gaussian, one obtains
\bea \label{quasi}
C_K({\bf r}) \sim r^{-\eta_K}
,\eea
with $\eta_K = \eta_K^{{\rm G}}:= \frac{1}{2} {\cal A}_2 K^2$, hence
quasi-long range translational order and sharp diffraction peaks, a characteristic of the BrG
\cite{GiamarchiLeDoussal1994,%GiamarchiLeDoussal1995,
KleinJoumardBlanchardMarcusCubittGiamarchiLeDoussal2001}.
This holds for space dimension $d_{\rm lc}<d<d_{\rm uc}$ (i.e.\ ${\bf r} \in \mathbb{R}^d$) with
$d_{\rm lc}=2$, $d_{\rm uc}=4$ for standard local elasticity. It
was obtained by variational methods
and confirmed by the Functional
renormalization group (FRG) \cite{Nattermann1990,GiamarchiLeDoussal1994
%,GiamarchiLeDoussal1995
}, a field-theoretic method  developed in recent years \cite{DSFisher1986,
%NattermannStepanowTangLeschhorn1992,
ChauveLeDoussalWiese2000a,LeDoussalWieseChauve2003,LeDoussal2006b,
%LeDoussal2008,
WieseLeDoussal2006,LeDoussalWiese2008c}, which
allows to treat multiple metastable states. The FRG predicts the universal
 amplitude ${\cal A}_2$ in a dimensional expansion in $d=d_{\rm uc}-\epsilon$. In this letter  we restrict for simplicity
 to the scalar case $N=1$, i.e.\ $u(\mathbf{r}) \in \mathbb{R}$, and
  choose the periodicity of $u$ to be one, hence the RLV
 to be $K=2 \pi k$ with $k$ integer. Then, within a 2-loop FRG calculation
 \cite{LeDoussalWieseChauve2003},
${\cal A}_2=\frac{\epsilon}{18} + \frac{\epsilon^2}{108} + {\cal O}(\epsilon^3)$
in agreement with numerics \cite{McNamaraMiddletonChen1999,NohRieger2001} for $d=3$.

The rationale for the Gaussian approximation is
that around $d_{\rm uc}$
%the statistics is nearly Gaussian, i.e.
one can decompose $u = \sqrt{\epsilon} u_1 + \epsilon u_2 +...$ into
independent fields $u_i$, where $u_1$ is Gaussian
(see Appendix G of \cite{LeDoussalWiese2008c}).
%\footnote{see Appendix G of \cite{LeDoussalWiese2008c}}.
% from a central limit theorem
Hence non-Gaussian corrections to
$\eta_K$ are expected only to ${\cal O}(\epsilon^4)$. However they
grow rapidly with $K$ and surely become important for secondary Bragg peaks.
This motivates a calculation of the higher cumulants of $u({\bf r})$. We also want to study
$C_K({\bf r})$ for arbitrary $K=2 \pi k$ with $k$ not necessary an integer.
This is needed e.g.\ in the context of the roughening transition \cite{EmigNattermann1997}
to determine whether the BrG is stable to a
small periodic perturbation $V_K= \int \rmd^d {\bf r}  \cos(K u({\bf r}))$.
Finally, for the algebraic decay (\ref{quasi}) to hold for all
$K$  all cumulants need to grow as $\ln r$,
a property which we demonstrate.

Another motivation to study the higher cumulants of $u({\bf r})$ comes from multifractal statistics, with examples ranging from turbulence \cite{Paladin}
to localization of quantum particles \cite{EversMirlin2008}.  Although $u({\bf r})$ exhibits
single-scale fractal statistics, we show here that the {\it exponential field} $e^{u({\bf r})}$
exhibits multifractal scaling, i.e.\ its moments behave with system size $L$ as
\bea \label{xq}
\overline{\langle e^{q u({\bf r}) }  \rangle} \sim \Big(\frac{a}{L}\Big)^{\!x_q}
,\eea
with a scaling dimension $x_q$.
%that we calculate.
This provides an interesting example beyond the well-studied Gaussian case
\cite{CastilloChamonFradkinGoldbartMudry1997,CarpentierLeDoussal2001}
of the general correspondence between exponentials of log-correlated fields and
statistically self-similar and homogeneous multifractal fields \cite{Fyodorov2010}.

The aim of this letter is thus to go beyond the Gaussian approximation: We
calculate the multifractal exponents $x_q$ and obtain
the higher cumulants of the log-correlated displacement field $u$ as
\be \label{cum1}
\overline{\left<[u({\bf r})-u(0)]^{n}\right>^{\rm c} } \simeq {\cal A}_{n}  \ln ({r}/{a})
\ee
for $r \gg a$, $n$ even, where each ${\cal A}_{n}$ is calculated to leading order in $\epsilon=4-d$ (odd cumulants
vanish by parity $u \to -u$). We use the FRG and develop a method based on
the asymptotic evaluation of functional determinants, which allows us to sum up
an {\em\ infinite subset of diagrams}. Amazingly, it can also be applied
to compute integrals appearing in a perturbative calculation
on the field-theory  side of AdS/CFT, known
as Konishi integrals \cite{EdenHeslopKorchemskySmirnovSokatchev2012}.

Let us mention that for the same model in $d=d_{\rm lc}=2$ (the Cardy-Ostlund model) such a summation was  achieved
using conformal perturbation theory
\cite{LeDoussalRistivojevicWiese2013}. While for $d>2$ the
${\cal A}_{n}$ are $T$ independent, in $d=2$ the glass phase
is marginal and exists for $T<T_{\rm c}$. The higher cumulants, as well as $C_{K}({\bf r})$ for $k \leq 1$,
were obtained to leading order in $T_{\rm c}-T$.

%%%%%%%%%%%%%%%%%%%%%%%%  Model %%%%%%%%%%%%%%%%%%%

{\it The model:} The Hamiltonian of an elastic system in a disordered environment can be written as
\begin{equation}
{\cal  H}[u]  =    \int_{\mathbf{x} }
 \frac{1}{2}[\nabla u(\mathbf{x})]^2 +\frac{m^2}{2} u^2(\mathbf x)+
 V(u(\mathbf{x}),\mathbf{x}), \label{eq-hamiltonian}
\end{equation}
%where we have introduced the elastic energy kernel
%$g_{\mathbf{r} \mathbf{r}'}=\int_q g_q \rme^{i q(\mathbf{r}-\mathbf{r}')}$, %and
%$\int_q \equiv \frac{\rmd^d q}{(2 \pi)^d}$. For
%short-ranged elasticity, it is given by  $g^{-1}_q = q^2 + m^2$.
with $\int_{\mathbf x}:= \int \rmd^d {\mathbf x}$. The first term is the elastic energy. The second term is a confining potential
with curvature \(m^2\) which effectively divides the system into independent subsystems of
size $L_m = 1/m$, hence provides an infrared (IR) cutoff. The random potential $V(u,\mathbf x)$ is a Gaussian  with zero mean and correlator
\begin{equation}  \label{bare}
\overline{V(u,\mathbf{x}) V(u',\mathbf{x}')}=R_0(u-u')
\delta^d(\mathbf{x}-\mathbf{x}'),
\end{equation}
where $R_0(u)$ is a function of period unity, reflecting the
periodicity of the unperturbed crystal \cite{GiamarchiLeDoussalBookYoung}.
The partition function in a given
disorder realization, at temperature $T$, is
 ${\cal Z}:=\int {\cal D}[u]\, \rme^{-{\cal H}[u]/T}$.
To average over the disorder,  we introduce replicas $u_\alpha(\mathbf x)$, $\alpha=1,\ldots,\sf n$ of
the original system. This leads to the bare replicated action \begin{eqnarray} \label{eq-action}
{\cal  S}_{R_0}[u] &=& \frac{1}{T} \sum_\alpha     \int_{{\mathbf x}}
 \frac{1}{2}[\nabla u_\alpha(\mathbf{x})]^2 +\frac{m^2}{2} u_\alpha^2(\mathbf x) \nn \\
&& - \frac{1}{2 T^2} \sum_{\alpha \beta} \int_{\mathbf x} R_0\big(u_\alpha(\mathbf{x})-u_\beta(\mathbf{x})\big).
\end{eqnarray}
The observables of the disordered model can be obtained from
those of the replicated theory in the limit ${\sf n}\to 0$.

{\it FRG basics:} The central object of the FRG is the renormalized disorder
correlator, the $m$-dependent function $R(u)$. Appropriately defined from the effective action $\Gamma[u]$ associated
to ${\cal  S}_{R_0}[u]$, the function $R(u)$ is an observable \cite{LeDoussal2006b
%,LeDoussal2008
},
which has been measured in numerics \cite{MiddletonLeDoussalWiese2006
%,RossoLeDoussalWiese2006a
} and
in experiments \cite{LeDoussalWieseMoulinetRolley2009}. It satisfies a FRG flow equation
as $m$ is decreased to zero ($R=R_0$ for $m=\infty$). Under rescaling,
$R(u)=A_d m^{\epsilon - 4 \zeta} \tilde R(m^\zeta u)$, with $A_d=
%\frac{1}{\epsilon \tilde I_2}=
\frac{(4 \pi)^{d/2}}{\epsilon \Gamma(\epsilon/2)}$,
%,$A_4=8 \pi^2$
$\tilde R(u)$
admits a periodic fixed point
(FP) with $\zeta=0$, and  $u \in [0,1]$,
%uniformly of order ${\cal O}(\epsilon)$
%:
\bea
\tilde R^*(u)- \tilde R^*(0) =  \tilde R^{* \prime \prime}(0) \frac{1}{2} u^2 (1-u)^2.
\eea
This form is valid for any $d<4$, and $- \tilde R^{* \prime \prime}(0) = \frac{\epsilon}{36} + \frac{\epsilon^2}{54}$
to two loop accuracy, in agreement with numerics \cite{MiddletonLeDoussalWiese2006}.
%Here .
The salient feature is that the renormalized force correlator $-R''(u)$ acquires a cusp at $u=0$, which we
denote by $\tilde \sigma = \tilde R^{*\prime \prime \prime}(0^+)= \frac{\epsilon}{6}+\frac{\epsilon^2}{9}$.
This cusp, seen in experiments \cite{LeDoussalWieseMoulinetRolley2009},
is the hallmark of the multiple metastable states
and is directly related to the statistics of shocks and avalanches which occur when
applying an external force \cite{LeDoussalWiese2008c}.

\begin{figure}
%\Fig{diagrams-letter}
\Fig{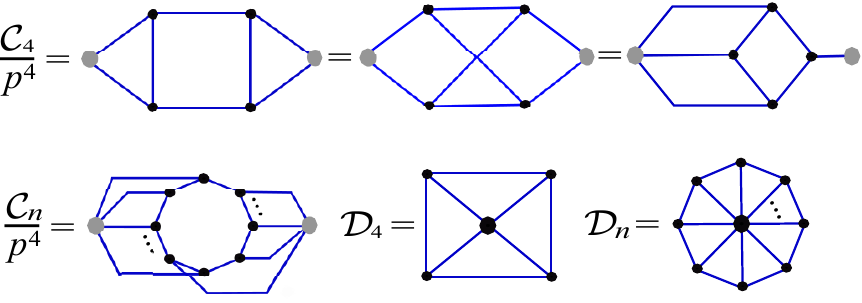}
\caption{Diagrammatic representation of the integrals
contributing to the translational correlation function
to leading order. The $C_n$ have two external points
(big circles, grey) where the external momentum $p$ enters. They are constructed from a polygon with $n$
vertices each attached to one of the two external points.
They are finite in $d=4$ and $\sim 1/p^4$. ${\cal D}_n$ has one external point (big circle, not
integrated over) all other points are integrated over. It is log-divergent in $d=4$.}
\label{fig-diag}
\end{figure}

{\it Determinant formula:} The cumulants (\ref{cum1}) can be computed
from (\ref{eq-action}) in perturbation theory  in $R_0$ at $T=0$, the leading order
being ${\cal O}(R_0'''(0^+)^n)$. This perturbation theory involves (complicated) replica combinatorics, see e.g.\ \cite{LeDoussalWieseChauve2003}. It also requires the
evaluation of multi-loop integrals represented
%diagrammatically
in fig.~\ref{fig-diag},
%an a-priori
a formidable task. We now show how to shortcut these difficulties.
We first reduce the problem to the calculation of a functional
determinant using the method developed in \cite{LeDoussalWiese2011b}
to evaluate averages of the form
$\mathcal{G}[\lambda]:=\overline{\left\langle{\exp\big( \int_{\mathbf{x}} \lambda(\mathbf{x}) u (\mathbf{x}) } \big) \right\rangle}=%
\lim\limits_{{\sf n}\to 0}\left< \exp\big({ \int_{\mathbf{x}} \lambda(\mathbf{x}) u_1(\mathbf{x}) } \big)\right>_{\cal S}$
where $u_1(\mathbf{x})$ stands for one of the $\sf n$ replicas.
The  function $C_{K}({\bf r})$ can  then be computed using  the charge density
of a dipole, $\lambda_{\rm D}(\mathbf{x}):=i K [\delta(\mathbf{x}-\mathbf{r})-\delta(\mathbf{x})]$.
For an arbitrary $\lambda(\mathbf{x})$, the average is expressed
%in terms of the effective action $\Gamma[u]$
as
$ \mathcal{G}[\lambda]  =
\exp( \int_{\mathbf{x}} \lambda(\mathbf{x}) u^\lambda(\mathbf{x}) - \Gamma[u^\lambda])$,
where $u^\lambda(\mathbf{x})$ extremizes the exponential,
i.e.\ is solution of
%the equation
$ \partial_{u_a(\mathbf{x})} \Gamma[u]\big|_{u=u^\lambda}%
= \lambda(\mathbf{x}) \delta_{a1}$. The effective action was
calculated in an expansion in
%powers of
$R$ (i.e.\ in $\epsilon$) to leading order
(one loop) as $\Gamma[u]={\cal  S}_R[u] + \Gamma_1[u] $
where ${\cal  S}_{R}[u]$ is the improved action with
the bare correlator $R_0$ replaced by the renormalized one $R$,
%, where $R_0$ is replaced by $R$ the renormalized correlator
and $\Gamma_1[u]$  is displayed e.g.\
 in\cite{ChauveLeDoussal2001,LeDoussalWiese2011b}.
Performing the extremization at $T=0$, a slight generalization
of section IV.A of Ref. \cite{LeDoussalWiese2011b}
leads to
\bea
&& \overline{\left\langle{e^{\int_{ \mathbf{x}} \lambda(\mathbf{x}) u (\mathbf{x})} } \right\rangle}
= \mathcal{G}_{{\rm Gauss}}[\lambda] e^{- \Gamma_\lambda} \label{eq9}
.\eea
Here $\mathcal{G}_{{\rm Gauss}}[\lambda] =e^{\frac{1}{2} \int_{{\bf x} {\bf x'}} \lambda({\bf x}) \lambda({\bf x'}) \overline{\langle u(\mathbf{x}) u(\mathbf{x}')\rangle} }$ is the Gaussian approximation,
 $\overline{\langle u({\bf x}) u({\bf x'}) \rangle }$ the exact 2-point correlation function, and
the effective action is\bea \label{fermionic-sign}
 - \Gamma_\lambda =
\frac12 \Big\{ \ln \mathcal{D}_{\rm reg}[\sigma U(\mathbf{r})]+\ln \mathcal{D}_{\rm reg}[-\sigma U(\mathbf{r})]
\Big\}.
\eea
The effective disorder is $\sigma:=R'''(0^+)$, and we define
\begin{equation} \label{eq-fun-det-1}
 \mathcal{D}[\sigma U(\mathbf{r})] :=
 \frac{\det ( - \nabla^2 +  \sigma U(\mathbf{r}) + m^2 ) }{
  \det ( - \nabla^2 + m^2 ) } .
\end{equation}
Its logarithm,  $\ln (\mathcal{D}[\pm\sigma
U],   $  has a perturbative expansion in \(\sigma\). The first two terms,
of order $\sigma$ and $\sigma^2$, which contain ultraviolet divergences in $d=4$,
are included in the Gaussian part. The remaining terms, i.e.\ all ${\cal O}(\sigma^p)$ with $p \geq 3$, define the regularized
determinant  $\ln (\mathcal{D}_{\rm reg}[\pm\sigma U])$.
%Its logarithm,   $\ln (\mathcal{D}[\pm\sigma
%U],   $  has a perturbative expansion in \(\sigma\), of which the first two terms, of order
%$\sigma$ and $\sigma^2$, are ultraviolet divergent; they are  included in the Gaussian part. The remaining terms  define the regularized
%determinant   $\ln (\mathcal{D}_{\rm reg}[\pm\sigma U])$.
Thus (\ref{fermionic-sign}) contains
only information about higher cumulants\footnote{A simpler version of (\ref{fermionic-sign}) was considered in Appendix G of \cite{LeDoussalWiese2008c}
for a uniform source; it yields the cumulants of $\int_{\bf r} u({\bf r})$.}. We have introduced the potential
\begin{equation}
U(\mathbf{r}) := \int_{{\bf x}} (-\nabla^2+m^2)^{-1}_{\mathbf r,\mathbf x}\,
\lambda(\mathbf{x}),
\end{equation}
which in the limit $m\to0$ satisfies the $d$-dimensional
Poisson equation $ \nabla^2 U(\mathbf{r}) = - \lambda(\mathbf{r})$.
%
%Note that straight perturbation theory leads
%to the same formula where $R$ is replaced in () and () by the
%bare disorder $R_0$ and without the subtraction $I_2$.
%This however leads to divergences which are automatically removed
%in the present, regularized form, since $R'''(0^+)$ is itself an observable.
%Note that the sign in eq.\ (\ref{fermionic-sign}) is opposite to the expectation for a standard, i.e.\ disorder-less, $1$-replica bosonic theory.
Note that two copies of the determinant appear in the present static
problem in eq.~(\ref{eq9}) as $\sqrt{  \mathcal{D}[\sigma U]   \mathcal{D}[- \sigma U] }$,
which can thus be interpreted as originating from an {\em effective fermionic}
field theory  with two flavors of real fermions.
A related observation was made in a dynamical calculation of
the distribution of pinning forces at the depinning transition
~\cite{%LeDoussalWiese2003a,
FedorenkoLeDoussalWiese2006}, where
only one copy appears, as $\mathcal{D}[\sigma U]$. Note
also, from fig.~\ref{fig-diag}, that to this order we have an
effective {\it cubic} field theory with coupling $\sigma$.
%Moreover, as we will see below, in presence of the symmetry
%$\lambda\to- \lambda$, both determinants are identical
%and the two  real fermions can be merged into a single  complex one. {\blue ***???***}
The 2-point correlation function in Fourier\footnote{It
was calculated to ${\cal O}(\epsilon^2)$ in
\cite{LeDoussalWieseChauve2003} Sec.\ VI A.}  reads
$\overline{\langle u_{p} u_{-p} \rangle } = c_d p^{-d} f(p/m)$,
with $f(z) \sim \tilde c_d z^d/c_d $ for small $z$,   $f(\infty)=1$,
%with $f(0)=1$ and $f(z) \sim \tilde c_d z^d/c_d $ at large $z$,
$\tilde c_d = - A_d \tilde R^{*\prime \prime}(0)$ and $c_d=\tilde c_d(1-\epsilon+...)$.
Inserting this with the 1-loop FP value into
$\mathcal{G}_{{\rm Gauss}}[\lambda]$
leads to the above Gaussian result for $\eta^{G}_K$
with
${\cal A}_2 = \frac{2 S_d c_d }{(2 \pi)^d}$, and
$S_d=\frac{2 \pi^{d/2}}{\Gamma(d/2)}$.

%Since ${\cal A}_2 = \frac{2 S_d c_d }{(2 \pi)^d}$, and
%$S_d=\frac{2 \pi^{d/2}}{\Gamma(d/2)}$, inserting the FP values,
%$\mathcal{G}_{{\rm Gauss}}[\lambda]$ leads
%to the above Gaussian result with $\eta^{G}_K$ .

%%%%%%%%%%%%%%%%%    Two-point function in annulus geometry  %%%%%%%%%%%
%%%%%%%%%%%%%%%%%%%%%%%%%%%%%%%%%%%%%%%%%%%%%%%%%%%%%%%%%%%%%%%%%%%%%%%%%%

{\it Evaluation of the determinant:}
We now have to evaluate the functional determinant (\ref{eq-fun-det-1}).
Unfortunately, there is no general method in $d>1$ for
a
%generic,
non-spherically-symmetric potential. However, as we show below,
it is sufficient to calculate the determinant for a spherically symmetric potential,
 and then apply a multi-fractal scaling analysis \cite{CatesDeutsch1987,DuplantierLudwig1991,Fyodorov2010}.
 Thus we start by computing the scaling dimension $x_{q}=x_{-q}$, as
 defined from (\ref{xq}). To this aim we  calculate
$\mathcal{G}[\lambda]$ for a (regularized) point-like charge
$\lambda_{\rm p}(\mathbf{r}):=q \delta_a(\mathbf{r})$
in a finite-size system. Since the corresponding potential is spherically symmetric,
to obtain the determinant
ratio (\ref{eq-fun-det-1})
we can employ
the Gel'fand-Yaglom method \cite{GelfandYaglom1960},
generalized to $d$ dimensions \cite{DunneKirsten2006}. We separate
the radial and angular parts of the eigenfunctions as
$\Psi(r,\vec{\theta})=\frac{1}{ r^{(d-1)/2}} \, \psi_{l}(r)\,%
Y_{l}(\vec{\theta})$, where the angular part is given by
a hyperspherical harmonic $Y_{l}(\vec{\theta})$,
labeled in part by a non-negative integer $l$.
The radial part $\psi_{l}(r)$ is an eigenfunction of the 1D (radial)
Schr\"odinger-like  operator ${\mathcal H}_{l} + \sigma U(r) +m^2 $,
where
\begin{equation}
{\mathcal H}_{l}:=
-\frac{\rmd^2}{\rmd r^2}+\frac{\left(l+\frac{d-3}{2}\right)
\left(l+\frac{d-1}{2}\right)}{r^2}.
\end{equation}
The logarithm of (\ref{eq-fun-det-1}) can be
written as a sum of the logarithms of
the 1D determinant ratios ${\cal B}_l$ for partial waves
weighted with the degeneracy of
%the number of modes at
angular momentum $l$,
\begin{equation}
\ln\left(\mathcal{D} [\sigma U] \right)
=\sum_{l=0}^\infty \frac{(2l+d-2)(l+d-3)!}{l!(d-2)!}
\ln {\cal B}_l. \label{eq-dunne}
\end{equation}
The  Gel'fand-Yaglom method gives the ratio of the
1D functional determinants for each partial wave $l$ as
\begin{equation}%\label{NK1}
{\cal B}_l:=\frac{\det \left[{\mathcal H}_{l}+\sigma U(r) + m^2\right]}{\det
\left[{\mathcal H}_{l}+m^2\right]}=
\frac{\psi_l(L)}{\tilde{\psi}_l(L)}. \label{eq-det-l}
\end{equation}
Here %$\tilde{\psi}_l(r)=r^{l+(d-1)/2}$ and
${\psi}_l(r)$ is the solution of
the initial-value problem for
%the equation
\begin{equation} \label{eq-Schodinger}
\left[{\mathcal H_l}+\sigma U(r)+m^2\right]{\psi}_l(r)=0,
\end{equation}
satisfying
$ {\psi}_{l}(r) \sim r^{l+(d-1)/2}$ for $r\to 0$. Equation (\ref{eq-det-l})
holds for the boundary conditions $u(|\mathbf{r}|=L)=0$,
taking the large-$L$ limit afterwards\footnote{To work directly in an infinite system,
the electric field must vanish fast enough. One can either use $m=0$ with a neutral charge configuration (dipole),
or $m>0$ (screening, exponential decay).}.
The function $\tilde \psi_l({r})$  solves (\ref{eq-Schodinger}) with the same small-$r$ behavior, but for $\sigma=0$.

We can now calculate
$\overline{\langle e^{q u(\mathbf{r})}\rangle}$ to leading order
in $d=4 - \epsilon$. Since $\sigma={\cal O}(\epsilon)$ we can
perform the calculation in $d=4$.
A point-like charge distribution leads to a potential
$U(r)\sim 1/r^{d-2} $ which is too singular at
the origin in $d=4$. % the functional determinant is
%well defined only for potentials which have singularity
%less stronger than $1/r^{2}$.
We introduce an UV cutoff via a uniformly charged  ball of radius $a$,
$\lambda_{\rm B}(\mathbf{r}) = \frac{q d}{S_d a^d}\Theta(a-|\mathbf{r}|) $.
Since $L$ is finite, we  solve Poisson's equation setting $m\to 0$ and obtain
\begin{equation}\label{eq-ball-pot}
U (r) = \left\{\begin{array}{cl}
\displaystyle \frac{ q a^{2-d}} {2S_d } \left(\frac{d}{d-2}-\frac{r^2}{a^2}\right) \quad& ~\mbox{for}~ 0<r<a,\\
\displaystyle \frac{ q\rule{0mm}{2ex}}{S_d(d-2)} \frac{1}{r^{d-2}}  \quad& ~\mbox{for}~ a<r<L. \\
\end{array} \right.
\end{equation}
We insert this potential in the Gaussian approximation
which reads $\ln \mathcal{G}_{{\rm Gauss}}= - \frac{1}{2} R''(0) \int_{\bf r} U(r)^2$, to lowest order ${\cal O}(\epsilon)$.
The log-divergence of this integral in $d=4$ leads to
$x_q^{\rm G}=  - \tilde c_4 q^2/(8S_4)=- \epsilon q^2/72$.
More generally, eq.\ (\ref{logc}) requires by consistency that
$\overline{u({\bf r})^2} \simeq \frac{1}{2} {\cal A}_2 \ln (L/a)$
hence $x_q^{\rm G} = - {\cal A}_2 q^2/4$, fixing the quadratic part ${\cal O}(q^2)$ of $x_q$.

To calculate the leading non-Gaussian corrections to $x_q$ via
(\ref{eq-fun-det-1}), we find the solution of (\ref{eq-Schodinger}) in $d=4$ with the potential
(\ref{eq-ball-pot}).
%and the small $r$ conditions discussed above.
It reads, for $r<a$ \begin{equation}
\psi _l(r)= \frac{r^{l+\frac{3}{2}}}{ e^{\frac{ i r^2 \sqrt{s}}{2 a^2}}}
 \, _1F_1\left(\frac{l +2 -i \sqrt{s}}2+1;l+2;\frac{i r^2
   \sqrt{s}}{a^2}\right),  \label{eq-sol-a-1}
\end{equation}
and for $a<r<L$,\begin{equation}
\psi _l(r) = c_1 r^{\frac{1}{2}-\sqrt{(l+1)^2+s}}+c_2
   r^{\sqrt{(l+1)^2+s}+\frac{1}{2}}
.\end{equation}
We introduced $s:=\sigma q/(2 S_d)$.
One can find $c_{1,2}$ by matching
%both solutions and its derivatives
at $r=a$.
Using eq.~(\ref{eq-det-l}) we obtain
the partial-wave determinant, which is universal  at large $L$,
\begin{equation}\label{15}
 \ln {\cal B}_l =\left[\sqrt{(l+1)^2+s}-(l+1) \right]  \ln (L/a)+\mathcal O(L^0).
\end{equation}
The term $\mathcal O(L^0)$ can be calculated from the $c_i$; it is
not universal. Note that the massive problem also leads to
(\ref{15}) with $\ln( L)$ replaced by $\ln (1/m$).

Substituting this result  into eq.~(\ref{eq-dunne}) yields the result for  \(\ln (\mathcal{D} [\sigma U] ) \). However, the sum over $l$ diverges, indicating that this functional determinant
%, by itself,
requires
regularization in $d\ge2$~\cite{DunneKirsten2006}. However in
(\ref{fermionic-sign}) we only need the regularized
determinant
%defined through
%$\ln (\mathcal{D}_{\rm reg}[\pm\sigma U]) = \ln (\mathcal{D}[\pm\sigma U])|_{2}$ where all
%terms of $O(\sigma)$ and $O(\sigma^2)$ are subtracted. Then
$\mathcal{D}_{{\rm reg}}[\pm\sigma U] \sim (L/a)^{-F_{ \rm reg}(\pm s)}$ where the first two orders in $s$ are subtracted,
%the regularized exponent is
\begin{eqnarray}\label{16}
 F_{ \rm reg}(s)&=&-\sum_{l=0}^{\infty} (l+1)^2 \left( \sqrt{(l+1)^2+s} -(l+1)\nn\right. \\
&&\left. -\frac{s}{2(l+1)}+\frac{s^2}{8 (l+1)^3}\right)\ .
\end{eqnarray}
Summing over $l$, it can also be written as a series in $s$,
\be \label{ser1}
F_{ \rm reg}(s) = \sum_{n=3}^\infty f_{n} s^n , \quad f_n =  (-1)^n \frac{\Gamma(n-\frac12) \zeta(2n-3)
  }{ 2 \sqrt{\pi}\Gamma(n+1) }.
\ee
Putting together the two copies we obtain the multi-fractal scaling exponent,  an even function of $s$ (and $q$),
\bea \label{resxq}
&& x_q= - \frac{1}{4} {\cal A}_2 q^2  +  F(s) \label{eq-exponent}, \quad  \quad s= \frac{\epsilon}{3} q, \\
&& F(s) := \frac{1}{2}\left[F_{\rm reg}(s)+F_{\rm
reg}(-s) \right] =  \sum_{n=2}^\infty f_{2 n} s^{2 n}
.\eea
To leading order we used $\sigma = A_d \tilde \sigma$, $\tilde \sigma = \frac{\epsilon}{6} +{\cal O}(\epsilon^2)$ and $S_4=2 \pi^2$.
%The symmetrized function appears due to the symmetry in eq.~(\ref{fermionic-sign}).
The final result is finite, as we avoided divergences by (i) using perturbation theory in the renormalized $R$
rather than in %\footnote{which leads to similar formula where %divergences must be subtracted via counter terms, leading to %the same final result}
the bare $R_0$,
(ii) by separating the non-Gaussian part $F(s)$ from the  Gaussian one. For completeness we also defined the single-copy exponent $F_{ \rm reg}(s)$ since it appears in the theory of depinning\footnote{At depinning, there is
an additional tadpole diagram associated to the non-zero average $\overline{u({\bf r})}=-F_c/m^2$, where $F_c$ is the threshold force. Similarly separating the non-Gaussian parts leads to $F_{ \rm reg}(s)$.}.

%%%%%%%%%%%%%%%%%%%%%%%%%%%%%%%%%%%%%%%%%%%%%%%
%\begin{widetext}
%\begin{figure*}
%\Fig{Freg}~~ \Fig{Fregsym}
%\caption{Numerical evalucation (dots) of the functions $F_{\rm reg} (s)$ (left) and $F_{\rm reg}^{\rm sym}
%(s)$ (right). We have also plotted the contribution of the mode $l=0$
%(red, solid), for the symmetrized function already almost
%indistinguishable from the exact result. For the non-symmetrized
%funtion we have also given the contribution of the $5$ first modes
%(green dashed), in excellent agreement with the full sum. }
%\label{f:F}
%\end{figure*}
%\end{widetext}
%%%%%%%%%%%%%%%%%%%%%%%%%%%%%%%%%%%%%%%%%%%%%%%%%
%%%%%%%%%%%%%%%%%%%%%%%%%%%%%%%%%%%%%%%%%%%%%%%
%\begin{widetext}
\begin{figure}
\Fig{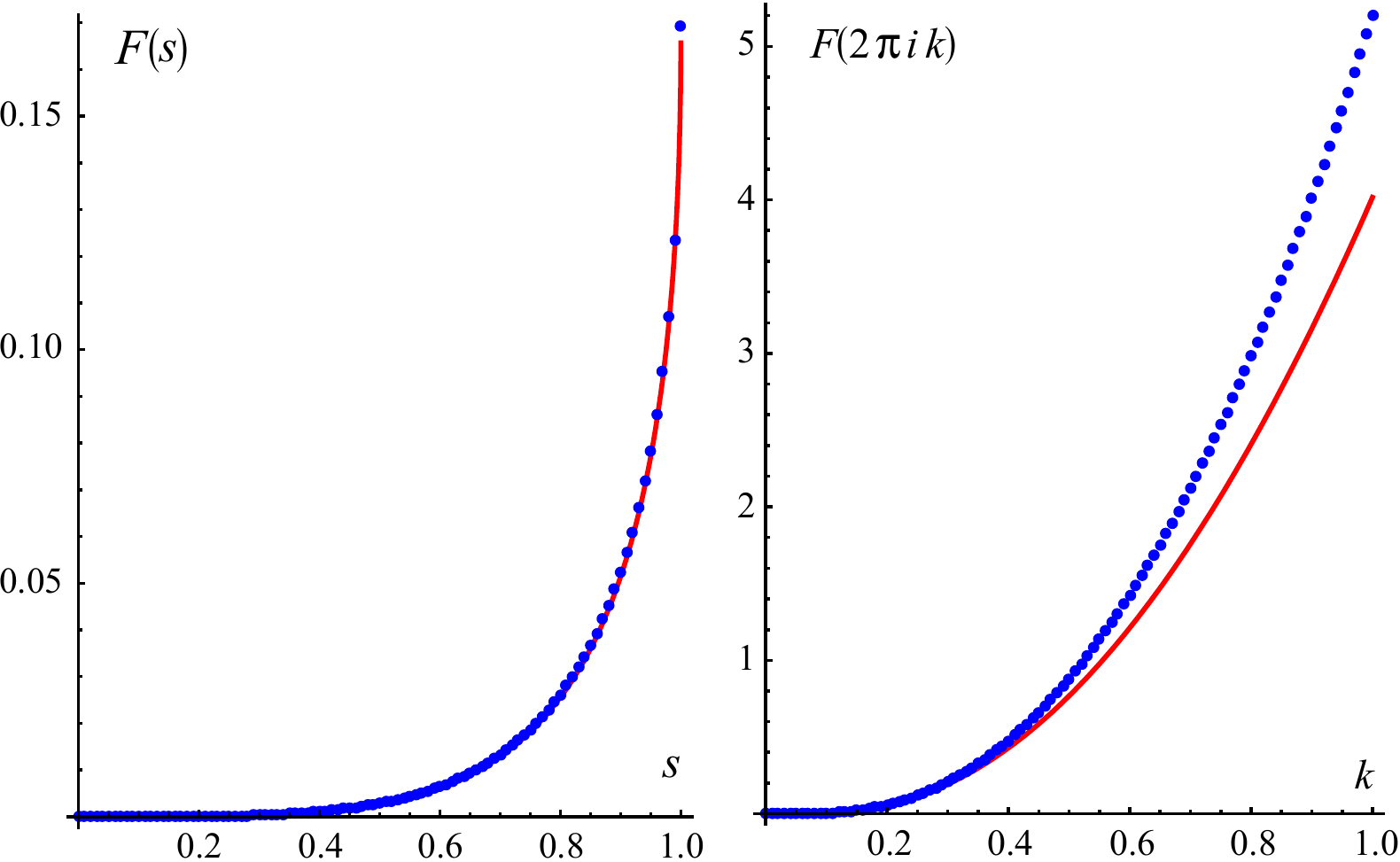}
\caption{Numerical evaluation (blue dots) of $F(s)$ (left) and $F(2\pi i k)$
(right). The red solid line is the contribution of
the mode $l=0$. }
\label{f:F}
\end{figure}
%\end{widetext}
%%%%%%%%%%%%%%%%%%%%%%%%%%%%%%%%%%%%%%%%%%%%%%%%%

{\it Analysis of the result:} Eq.\ (\ref{resxq}) is an even series in $s$ with a
radius of convergence of $|s|=1$. At $s=\pm 1$, $F(s)$,
plotted in fig.~\ref{f:F}, has a square-root singularity given by its $l=0$ term.
%The remainder $l \geq 1$ (which equals $0.00148536$ at $s=1$)
%has a radius of convergence $|s|=4$ and so on.
On the other hand, the exponent $x_q$ must satisfy\footnote{Since $\overline{\langle q u \sinh q u \rangle} \geq 0$ and
from Cauchy-Schwarz the inequality
$\overline{\langle u^2 e^{q u} \rangle} ~  \overline{\langle e^{q u} \rangle} \geq \overline{\langle u e^{q u} \rangle}^2$ must hold.}  $q \frac{\rmd}{\rmd q} x_q  \leq 0,$ and
convexity $\frac{\rmd^2}{\rmd q^2} x_q \leq 0$,
both requirements for multifractal field theories \cite{DuplantierLudwig1991}.
While the Gaussian part
$x^G_q= - \frac{1}{4} {\cal A}_2 q^2$ does, the correction term
$F(s)$ does not, since $F''(s)  \ge 0$.
Since $F''(s) \sim \frac{1}{8 (1-|s|)^{3/2}}$ diverges at $s = \pm 1$ ($q= q_p \simeq \frac{3}{\epsilon}$) one cannot trust the calculation in that region\footnote{Our result is a summation of a convergent
series in $q \epsilon$, but there is no guarantee that there are no non-perturbative corrections.}; it surely fails when $F''(\frac{q
\epsilon}{3}) > \frac{1}{4 \epsilon}$.

{\it Calculation of 2-point correlations:} To obtain the cumulants (\ref{cum1}) and the translational correlation function
(\ref{quasi})
%(via analytical continuation $q \to i K$)
we would need a dipole source, for which we cannot solve the Schr\"odinger problem. One way to proceed is to {\em assume} that the exponential field $e^{u({\bf r})}$ obeys
the conventional multifractal scaling formula \cite{CatesDeutsch1987,DuplantierLudwig1991,Fyodorov2010}:
\begin{equation}\label{eq-mult-scal}
\overline{\langle e^{q_1 u(\mathbf{r}_1) } e^{q_2 u(\mathbf{r}_2)} \rangle}
\sim  \Big(\frac{r_{12}}{a}\Big)^{x_{q_1+q_2}-x_{q_1} -
x_{q_2}} \Big(\frac{L}{a}\Big)^{- x_{q_1+q_2}}
,\end{equation}
with $r_{12}=|{\bf r}_1-{\bf r}_2|$. Since we already calculated $x_q$, this formula, taken for $q_1=-q_{2}=q$
immediately yields
\begin{equation}
   \overline{\left< e^{q [u(\mathbf{r}) - u(0)]}\right>}
 \sim \Big(\frac{r}{a}\Big)^{-2 x_{q}} \label{eq-transl-corr-res},
\end{equation}
using that $x_{q}=x_{-q}$ and $x_0=0$. Let us define the expansion
$x_q = \sum_{n=1}^\infty \frac{1}{n!} a_n q^n$. Using the standard formula
\bea \label{stand}
\ln \overline{\langle e^A \rangle}=\sum_{n=1}^\infty \frac{1}{n!} \overline{\langle A^n \rangle}^c
,\eea
we obtain one of the main results of this letter, eq.\ (\ref{cum1}), with the amplitudes for even $n \geq 4,$
\bea \label{main1}
{\cal A}_{n} = - 2 a_{n} = - \frac{\Gamma(n - \frac{1}{2})
\zeta(2 n -3)}{\sqrt{\pi}} \Big(\frac{\epsilon}{3}\Big)^{n}
.\eea
 There is actually more information in eq.\
(\ref{eq-mult-scal}): Using (\ref{stand}) and expanding in powers of $q_1^j q_2^{n-j}$ we obtain\bea \label{miracle1}
&& \overline{\langle u(\mathbf{r}_1)^j u(\mathbf{r}_2)^{n-j} \rangle}^c \simeq a_n \ln (r_{12}/L), \\
&& \overline{\langle u(\mathbf{r}_1)^n \rangle}^c \simeq - a_n \ln (L/a). \label{mn}
\eea
While we already know (\ref{mn}) from (\ref{xq}) and (\ref{stand}), eq.\ ({\ref{miracle1}),
valid for any $1 \leq j \leq n-1$ represents strong constraints.

Formula (\ref{eq-mult-scal}) is, at this stage, an {\em educated guess}, since we do not know the exact solution to the corresponding 2-charge
(dipole) Schr\"odinger problem.
%Here however we have a first-principle way to {\it calculate} %the l.h.s in (\ref{eq-mult-scal}) through our functional
%determinant by choosing the appropriate source.
%Although ,
We now close this gap via a careful examination of the integrals appearing in the expansion of
the determinant in powers of $\sigma$, represented by the diagrams
in fig.~\ref{fig-diag}. We show two properties:

(i) All terms of the form eq.\ (\ref{miracle1}) are equal, and
independent of $j$: This {\it proves } that both eqs.~(\ref{eq-mult-scal})
and~(\ref{eq-transl-corr-res})  hold.

(ii) The topologically distinct integrals with the same $j$ are also all equal. This  remarkable property  goes beyond
what is needed for eq.~(\ref{miracle1}), and  provides simple expressions for  such integrals; as announced in the introduction, they are of interest in the AdS/CFT context.

For clarity, let us detail the term $n=4$ (setting $m=0$). The calculation of $\overline{\langle u({\bf r}_1)^2 u({\bf r}_2)^2 \rangle}$
involves two 3-loop integrals, $I_{\{2,2\}_1}(p)$ and $I_{\{2,2\}_2}(p)$, which are  represented by the  first two (topologically distinct) diagrams in fig.~\ref{fig-diag}.
The first is {\it equal} to the integral, with entering momentum $p,$
 $I_{\{2,2\}_1}(p):= \int_{\bf q} \frac{I({\bf p},{\bf q})^2}{q^2 ({\bf p}-{\bf q})^2}$
with $I({\bf p},{\bf q}):=\int_{\bf k} \frac{1}{k^2 ({\bf k}+{\bf p})^2 ({\bf k}+{\bf q})^2}$, $\int_{\bf q}
:= \int \frac{\rmd^d \bf q}{(2 \pi)^d}$.
The third diagram (i.e integral) is the only one entering in
the calculation of $\overline{\langle u({\bf r}_1)^3 u({\bf r}_2) \rangle}$. By power counting, these integrals
are {\it both UV and IR finite} in $d=4$, and scale as $ p^{-4}$;  we now determine their amplitude.
%We now prove that eq.~(\ref{eq-transl-corr-res}) sums all diagrams of the type
%shown in fig.~\ref{fig-diag}.

First we show that, for given $n$, the diagrams with two external points depicted in fig.~\ref{fig-diag} are {\it independent on how these points are attached to
the polygon vertices}. In a nutshell this is because they all scale as $p^{-4}$, and if we identify the two external points, we obtain {\it the same} integral ${\cal D}_n$ in fig.~\ref{fig-diag}. Explicitly, for $m=0$ and $d=4$, any of these diagrams has $n-1$ loops and
$2 n$ propagators, and reads
\begin{equation} \label{type1}
\parbox{2.2cm}{\fig{2.2cm}{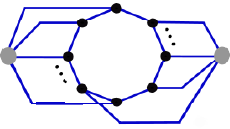}} =\frac{\mathcal{C}_n}{p^{4}}
,\end{equation}
where {\it a priori} $\mathcal{C}_n$ depends on how
we attach the $n$ points of the polygon to the two external points. In a massive scheme, and $d=4-\epsilon$, by power counting this changes to \begin{equation}\label{22}
\parbox{2.2cm}{\fig{2.2cm}{Cn}}=
\frac{\mathcal{C}_n}{p^{4+(n-1)\varepsilon}}\, g_n\!\left(\frac{p}{\alpha_n
m}\right),
\end{equation}
where $g_n(x)\to 1$ for $x\to \infty$, $g_n(0)=0$ and $\alpha_n$
parameterizes the crossover point with $g_n(1)=\frac{1}{2}$.
%Since in $d=4$ the integral $\int_{p}1/p^{4}$ is logarithmically divergent, we can either calculate the diagram ${\cal D}_n$ in $d=4$ using both an UV and an IR regulator, or in $d=4-\epsilon$ dimensions, with only an IR regulator.
%We use a mass $m$, keeping in mind that the leading divergence is universal.
Now ${\cal D}_n$ is obtained from  ${\cal C}_n$ by integrating over the external momentum:
\begin{eqnarray}
{\cal D}_n&=&\int_{\bf p}
\frac{\mathcal{C}_n}{p^{4+(n-1)\varepsilon}}\, g_n\!\left(\frac{p}{\alpha_n m}\right)  \simeq
\mathcal{C}_n \frac{S_d}{(2 \pi)^d}  \int_{\alpha_n m}^\infty \frac{\rmd p}{p^{1+ n \epsilon}}
\nn\\
&=&  \frac{\mathcal{C}_n (\alpha_n m)^{-n\varepsilon} }{8 \pi ^2 n\varepsilon }
+{\cal O}\left(\epsilon ^0\right) =  \frac{\mathcal{C}_n  m^{-n\varepsilon} }{8 \pi ^2 n\varepsilon
}
+{\cal O}\left(\epsilon ^0\right).\label{eq-D-n} \ \ \ \ \
\end{eqnarray}
The leading pole in $\epsilon$ does not depend on $\alpha_n$, and is universal.
Since all these diagrams lead to the same value of ${\cal D}_{n}$, all integrals of the type (\ref{type1})
are {\it equal}, and in $d=4$ equal to ${\cal C}_{n}/p^{4}$.

We already know the integral ${\cal D}_{n}$ in $d=4$ from eqs.\  ({\ref{16}) and (\ref{ser1}),
by matching powers of $q$ in the expansion of the determinant with a point source,
$\ln  \mathcal{D}[\sigma U] = \sum_{n=1}^{\infty} \frac{(-1)^{n+1}}{n} {\cal D}_n (q\sigma)^n$
which yields ${\cal D}_n \simeq (-1)^n n f_n/(2 \pi)^{2n} \ln( \frac{L}{a})$ for any $n \geq 3$.
Interestingly, the Yaglom-Gelfand method  allows us to calculate
${\cal D}_n$ directly in $d=4-\epsilon$. For $d<4$ we can set  $a=0$ in the potential (\ref{eq-ball-pot}). The corresponding radial Schr\"odinger problem can be solved {\it exactly} as
\be
\psi_l(r) = r^{l+\frac{d-1}{2}} z_l(r) , \ \ z_l(r)=\, _0F_1\left(\frac{2 (l+1)}{\epsilon };\frac{2 s r^{\epsilon }}{(2-\epsilon) \epsilon
   ^2}\right). \nn \label{eq42}
\ee
Using the identity $\lim_{\epsilon \to 0} \epsilon \ln _0F_1(\frac{2(l+1)}{\epsilon}, \frac{\tilde s}{\epsilon^2})
= \sum_{n=1}^\infty  \frac{(-1)^{n+1} \Gamma(n-\frac{1}{2}) \tilde s^n}{2 n \sqrt{\pi} \Gamma(n+1) (l+1)^{2n-1}}$
we calculate to leading order in $\epsilon$, $\ln \mathcal{D}[\sigma U] \simeq \sum_{l=0}^\infty (l+1)^2\,   \ln z_l(L)$.
This yields the polygon integrals for $n \geq 3$ in the massive scheme,
\begin{equation}\label{eq-D-n-2}
{\cal D}_n= \parbox{1.2cm}{\fig{1.2cm}{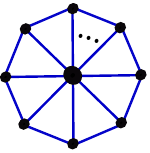}} = \frac{m^{-n\varepsilon}}{n \varepsilon }
      \frac{ \Gamma(n-1/2) \zeta(2n-3)
 }{2  \sqrt{\pi} (2\pi)^{2n}  \Gamma (n)} +{\cal O}(\epsilon^{0}).
\end{equation}
Note that $\frac{L^{n \epsilon}}{n \epsilon}$ changed to  $\frac{m^{-n \epsilon}}{n \epsilon}$. Further
substituting this factor by $\ln(L/a)$ reproduces the above estimate for $d=4$.

Using eqs.~(\ref{eq-D-n}) and (\ref{eq-D-n-2}) we now obtain $\mathcal{C}_n$ in $d=4$,
\begin{equation}\label{Cn}
{\cal C}_{n}= p^{4} \parbox{2.2cm}{\fig{2.2cm}{Cn}} = \frac{ \Gamma(n-\frac12) \zeta(2n-3)
  }{  \sqrt{\pi} \Gamma (n)(2\pi)^{2n-2}} .
\end{equation}
This allows to expand the determinant in presence of two charges $q_1$, $q_2$, in terms of 2-point diagrams,
and obtain, using (\ref{stand}) and (\ref{fermionic-sign}) in $d=4$ with $m=0$:
\bea
&& \sum_{n \geq 4} \frac{1}{n!} \overline{\langle [q_1 u({\bf r}) + q_2 u(0)]^n \rangle}^c
=\sum_{n ~ {\rm even} \geq 4 } \frac{(-1)^{n+1}}{n} \sigma^n  \nn \\
&& \times \bigg[ (q_1^n + q_2^n) {\cal D}_n + \int_{\bf p} e^{i  {\bf p} \cdot {\bf r} } {\sum_{j=1}^{n -1} } \left({n \atop j}\right) q_1^j q_2^{n-j} \frac{{\cal C}_n}{p^4} \bigg]
.\eea
Here we used that all ${\cal C}_n$ integrals are the same. Since $({n \atop j}) $ appears
on both sides it implies (\ref{miracle1}) with $a_n=-\frac{S_4}{(2 \pi)^4} {\cal C}_n (n-1)! \sigma^n$
in agreement with (\ref{main1}). Choosing $q_2=-q_1$  rederives our main result
for the cumulants (\ref{cum1}) and (\ref{main1}) since ${\sum_{j=1}^{n -1}} ({n \atop j})   (-1)^j=-2$.
We thus proved that the multifractal scaling relations (\ref{eq-mult-scal}) and (\ref{eq-transl-corr-res}) hold.

Performing the analytical continuation $q= i K$ we obtain the decay exponent\footnote{Note that $e^{ i K u(r)}$ obeys ordinary field-theory scaling, while
$e^{q u(r)}$ obeys multifractal scaling\ \cite{DuplantierLudwig1991}.} of the translational correlations,
\begin{equation}
\eta_{K} = \Big[\frac{\varepsilon}{36} + \frac{\varepsilon^2}{216} + {\cal O}(\varepsilon^3)\Big] K^2 + 2 F\Big(i K \frac{\varepsilon}{3}\Big)
.\end{equation}
The wave vector $K$ is arbitrary,   not necessarily a RLV\footnote{In $d=2$, $C_{K}(r)$ was argued  \cite{TonerDiVincenzo1990} to exhibit cusps for integer $K/(2 \pi)$
due to screening of the 2-point function by the interaction.}. Although
non-Gaussian corrections start at ${\cal O}(\epsilon^4)$, setting directly
$\epsilon=1$ and $K=K_0 = 2 \pi$
yields{\footnote{We used eq.~(\ref{16}) which can be considered as the analytic
continuation of eq.~(\ref{ser1}), whose radius
of convergence is $K=3$. }
$\eta_{K_0}^{\rm G}|_{\rm 1{\text{-}}loop}=1.097$, $\eta_{K_0}^{\rm G}|_{\rm2\text{-}loop}=1.279$
while $\eta_{K_0}-\eta_{K_0}^{\rm G}=0.569$.
Even if these
corrections may be an overestimate, and
  higher-loop corrections are needed,
%\footnote{\red A: I would erase this: The
% higher loop corrections are needed since to 1-loop the radius
%of convergence is $K=3$.}
non-Gaussian effects}\footnote{In $d=4$ the second cumulant grows as $\ln (\ln (r))$, while
higher ones reach a (non-universal) finite limit.} appear
to be non-negligible for $d=3$ \cite{NohRieger2001}.
Comparison with the elastic term \cite{EmigNattermann1997} then shows that a small periodic perturbation $V_K$ becomes relevant for $K<K_c$ with $2 - \eta_{K_c} =0$.

{\it Conclusion:} Using functional determinants we obtained the scaling
exponents of the (real and imaginary) exponential correlations of the displacement field
in a disordered elastic system. We leave calculating the spectrum of fractal dimensions\footnote{The Gibbs measure of a particle diffusing on top of
the elastic object with potential energy $\sim u(\mathbf{r})$
provides a normalized multifractal measure $\mu(\mathbf{r})=\frac{e^{\gamma u(\mathbf{r})}}{\int_{\mathbf x} e^{\gamma u(\mathbf{x})}}$ from which one can calculate a spectrum of dimensions.}, and the extension to a more general elastic kernels
for
the future.
%\cite{uslong}.
As a surprising corollary, our method yields, in an elegant way and for  arbitrary $n$,   exact expressions for the integrals
$\mathcal{C}_n$; (we  numerically checked formula (\ref{Cn}) for $n=3,4,5$).
Similar integrals appear in $N=4$ SYM, on the field-theory side of two theories related via
 AdS/CFT: E.g., $\mathcal{C}_5$ contributes to the Konishi anomalous dimension in $N=4$ SYM at five-loop order,
 and an elaborate formalism was put in place to calculate it \cite{EdenHeslopKorchemskySmirnovSokatchev2012}.
We hope that our method, and possible generalizations, will also allow for a further-reaching
check of the AdS/CFT duality\footnote{Reciprocally, the results in \cite{diagrams} yield the
full 4-point function for the Bragg glass.}.

\acknowledgements
We thank V.\ Kazakov and Y.\ Fyodorov for stimulating discussions.
This work was supported by ANR Grant 09-BLAN-0097-01/2.

\end{document}